\documentclass[reprint,amsmath,amssymb,aps]{revtex4-1}

\usepackage{graphicx}
\usepackage{dcolumn}
\usepackage{bm}
\usepackage{amsmath}
\usepackage{multirow}
\usepackage{float}
\usepackage{color}
\usepackage{ulem}
\usepackage{amssymb}
\usepackage{verbatim}

\begin{document}

\title{Temperature dependent anisotropy and two-band superconductivity revealed by lower critical field in organic superconductor $\kappa$-(BEDT-TTF)$_{2}$Cu[N(CN)$_{2}$]Br}

\author{Huijing Mu$^{1} $, Jin Si$^{1} $, Qingui Yang$^2$, Ying Xiang$^{1}$, Haipeng Yang$^{2,} $}
 \email{yanghp@szu.edu.cn}
\author{Hai-Hu Wen$^{1,} $}
 \email{hhwen@nju.edu.cn}
\affiliation{$^1$ National Laboratory of Solid State Microstructures and Department of Physics, Collaborative Innovation Center of Advanced Microstructures, Nanjing University, Nanjing 210093, China}
\affiliation{$^2$ College of Materials Science and Engineering, Shenzhen Key Laboratory of Polymer Science and Technology, Shenzhen University, Shenzhen, 518060, China}

\begin{abstract}
Resistivity and magnetization have been measured at different temperatures and magnetic fields in organic superconductors $\kappa$-(BEDT-TTF)$_{2}$Cu[N(CN)$_{2}$]Br. The lower critical field and upper critical field are determined, which allow to depict a complete phase diagram. Through the comparison between the upper critical fields with magnetic field perpendicular and parallel to the conducting ac-planes, and the scaling of the in-plane resistivity with field along different directions, we found that the anisotropy $\Gamma$ is strongly temperature dependent. It is found that $\Gamma$ is quite large (above 20) near $T_{c}$, which satisfies the 2D model, but approaches a small value in the low-temperature region. The 2D-Tinkham model can also be used to fit the data at high temperatures. This is explained as a crossover from the orbital depairing mechanism in high-temperature and low-field region to the paramagnetic depairing mechanism in the high-field and low-temperature region. The temperature dependence of lower critical field $H_{c1} (T)$ shows a concave shape in wide temperature region. It is found that neither a single $d$-wave nor a single $s$-wave gap can fit the $H_{c1} (T)$, however a two-gap model containing an $s$-wave and a $d$-wave can fit the data rather well, suggesting two-band superconductivity and an unconventional pairing mechanism in this organic superconductor.
\begin{description}
\item[PACS number(s)]
74.72.-h; 74.62.Bf; 62.50.-p
\end{description}
\end{abstract}

\maketitle

\section{\label{sec:level1}INTRODUCTION}
The series of $\kappa$-(BEDT-TTF)$_{2}$X is the most intensively studied organic superconductor family because of the relatively high superconducting transition temperature $T_{c}$ (above 10 K), low dimensionality, narrow bandwidth, large Coulomb correlation between conducting electrons, easily modulated ground states, etc. In these organic salts, BEDT-TTF stands for the bis(ethylenedithio)-tetrathiafulvalene (abbreviated as ET) and X$^{-}$  is a monovalent anion. These ET-based superconductors consist of alternative stacking of conducting ET layers and insulating anion layers. Within the conducting layers, every two ET molecules form a dimer, and the dimers form a triangular lattice in the $\kappa$-type structure. Since each dimer donates one electron to the anion layer, thus it possesses one hole, and the conduction band is half-filled\cite{kino1996phase}, which looks quite similar to the parent phase of cuprate superconductors. This layered crystal structure of $\kappa$-(ET)$_{2}$X leads to a high anisotropy between the stacking direction and the conducting planes direction, which consequently results in the high anisotropy of resistivity\cite{buravov1992anisotropic, sato1991electrical} and magnetic susceptibility\cite{PhysRevB.65.060505, PhysRevB.66.174521, PhysRevB.52.471, PhysRevB.49.15227}.

It is widely known that the high-temperature cuprate superconductors also have layered crystal structures with CuO$_2$ as the conducting planes. The $\kappa$-(ET)$_{2}$X salts and cuprate superconductors have many common features of physical properties, such as the coexistence and competition between the superconducting phase and other electronic states. The conceptual phase diagram of $\kappa$-(ET)$_{2}$X is qualitatively similar to that of the cuprate superconductors, with chemical pressure for the organic superconductors equivalent to carrier doping for the cuprate superconductors. The ground state of $\kappa$-(ET)$_{2}$X can be switched between the antiferromagnetic insulating, superconducting and metallic states by changing the anion X$^{-}$ in order to apply a chemical pressure. For example, $\kappa$-(ET)$_{2}$Cu[N(CN)$_{2}$]Cl (abbreviated as $\kappa$-(ET)$_{2}$ACl, with A = Cu[N(CN)$_{2}$]) is an antiferromagnetic insulator, but it shows superconductivity with $T_{c}$ of about 13 K under a moderate pressure of 0.3 kbar\cite{PhysRevB.44.4666, wang1992phase}. The isomorphic salt $\kappa$-(ET)$_{2}$Cu[N(CN)$_{2}$]Br (abbreviated as $\kappa$-(ET)$_{2}$ABr) is located on the metallic side of the Mott-Hubbard transition and shows superconductivity with $T_{c}$ of about 12 K at ambient pressure\cite{wang1991new}. Since the phase diagram of this organic superconductor $\kappa$-(ET)$_{2}$X system has lower temperature and pressure scales compared with that of cuprate superconductors, it is thus more easy to manipulate the ground state properties and depict the complete phase diagram. In addition, enormous efforts were dedicated to the possible interesting Fulde-Ferrell-Larkin-Ovchinnikov (FFLO) state\cite{PhysRev.135.A550, larkin1965inhomogeneous} in low-temperature and high-field region of $\kappa$-(ET)$_{2}$X system\cite{J.Singleton_2000, PhysRevB.85.214514, Fortune_2018, mayaffre2014evidence, wosnitza2018fflo}. The existence of a FFLO state is reflected by the upturn of upper critical field $H_{c2}$ in the low-temperature and high-field region, and the value of $H_{c2}(0)$ exceeds the Pauli-limiting field $H_{p}$\cite{imajo2021fflo}.

Elucidating the pairing mechanism plays an important role in understanding the possible origin of superconductivity. In the unconventional superconductivity of cuprate superconductors, the $d$-wave symmetry of order parameter has reached a broad consensus\cite{RevModPhys.72.969}. But the pairing mechanism of $\kappa$-(ET)$_{2}$X is still controversial. The $^{13}$C nuclear magnetic resonance (NMR) experiments and the Knight shift of $\kappa$-(ET)$_{2}$X suggest an unconventional pairing state with possible a nodal gap\cite{PhysRevLett.75.4122, de1995c}. The obvious fourfold oscillation of the magnetic field angle dependence of heat capacity provided strong evidence to support the $d$-wave symmetry\cite{PhysRevB.82.014522}. The $d$-wave symmetry is also supported by experiment data of scanning tunneling microscopy (STM)\cite{doi:10.1143/JPSJ.77.114707}, microwave surface impedance\cite{PhysRevB.88.064501} and angle dependent magnetothermal conductivity measurements\cite{PhysRevLett.88.027002}. However, there is also experimental evidence to support fully gapped order parameter, such as the exponential temperature dependence of electronic specific heat in the superconducting state\cite{PhysRevLett.84.6098, PhysRevB.65.140509}. In addition to the assumption of a single-gap model, there are theoretical calculations and experimental evidence for a two-gap model that includes an $s$-wave component and a $d$-wave component\cite{PhysRevB.97.014530, PhysRevLett.116.237001, refId0}, however, it is quite challenging to illustrate this combination of two gaps. The gap message, although still under debate, can be found in some experiments, like the angle resolved heat capacity\cite{PhysRevB.82.014522}, or some overview papers on such materials\cite{SingletonReview,WosnitzaReview}.

In this paper, we report results of resistivity and magnetization measurements of the organic superconductor $\kappa$-(ET)$_{2}$ABr. The anisotropy of $\kappa$-(ET)$_{2}$ABr were studied in detail by measuring the in-plane resistivity at different temperatures and magnetic fields with the field applied perpendicular and parallel to the conduction planes.  Two theoretical models, two dimensional (2D)-Tinkham model and anisotropic-3D-Grinzburg-Landau (GL) model were used to fit the angular dependence of $T_{c}(H)$ extracted from the experimental data, and it was found that the data is more suitable for the 2D-Tinkham model, indicating a strong 2D nature. The anistropy is thus determined and found to be strongly temperature dependent. In order to determine the lower critical field $H_{c1}$, we carried out a systematic measurement on the magnetization-hysteresis-loops of $\kappa$-(ET)$_{2}$ABr in the field penetration process, and the temperature dependence of $H_{c1}$ was well described by a two-gap model with one $s$-wave component and one $d$-wave component. By using the well determined data of $H_{c1}$, the magnetic field penetration depth $\lambda$ in low-temperature region was further derived.

\section{EXPERIMENTAL METHOD}
Single crystals of $\kappa$-(ET)$_{2}$ABr were grown by the electrochemical method\cite{doi:10.1021/ic00339a004, doi:10.1021/cm00011a003}. The electrolyte was the mixture of BEDT-TTF (98.0$\%$, TCI), CuBr (99.39$\%$, bidepharm),  NaN(CN)$_{2}$ (98$\%$, HEOWNS) and 15-crown-5 (97$\%$, Energy Chemical). The solvent was 1,1,2-trichloroethane (TCE) (99$\%$, MERYER). An appropriate amount of ethanol and deionized water were added to increase the conductivity of the solution. Plate-like single crystals were grown on Pt anode under galvanostatic condition (0.8 $\mu$A) at room temperature in a N$_{2} $ gas atmosphere for 30-40 days. The crystal sizes used for the resistivity measurement and magnetization measurement were $0.52\times0.52\times0.14$ mm$^{3} $ and $0.62\times0.56\times0.22$ mm$^{3} $, respectively. The temperature dependence of in-plane resistivity was measured by using the standard four-probe method in a physical property measurement system (PPMS, Quantum Design) with the magnetic fields up to 9 T. The angular dependence of in-plane resistivity was measured with the angle $\theta$ ranging from $0^{\circ} $ to $180^{\circ} $, where $\theta =0^{\circ} $ and $\theta =180^{\circ} $ correspond to the magnetic field perpendicular to the ac-planes of the single crystal, and $\theta =90^{\circ} $ corresponds to the case of magnetic field applied parallel to the ac-planes. The definition of the angle $\theta$ and the direction of the applied current are shown in the inset of Fig. \ref{fig2}(b). The dc magnetization was measured by a superconducting quantum interference device-vibrating sample magnetometer (SQUID-VSM, Quantum Design) with the magnetic fields up to 7 T, and the lowest measured temperature is 1.8 K. The stable mode was used to measure the low-field magnetization in order to obtain the lower critical field and penetration depth.

\section{Results and discussion}
\subsection{Characterization of the superconducting transition}
\begin{figure}
\centering
\includegraphics[width=8cm]{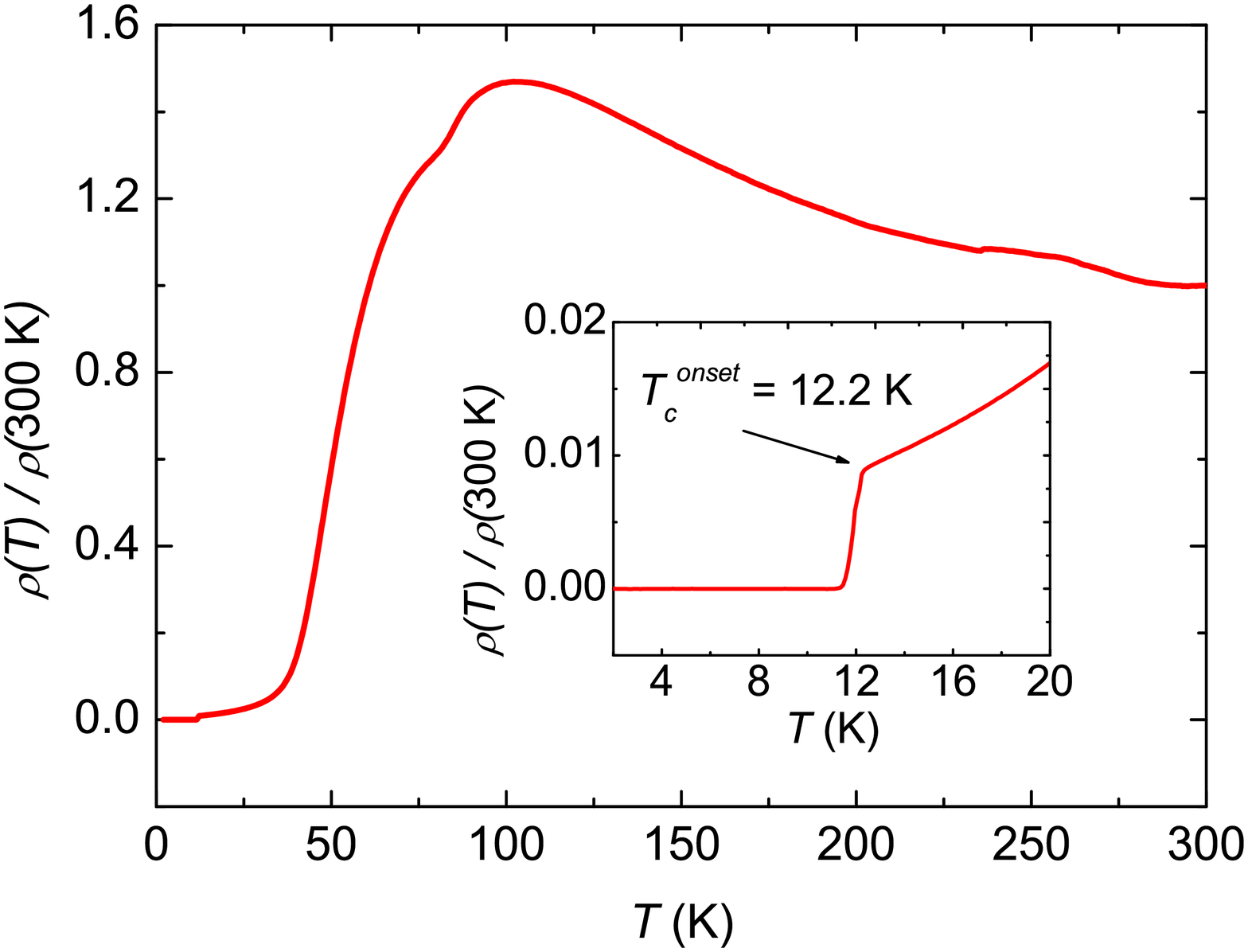}\\[5pt]
\caption{\label{fig1} Normalized in-plane resistivity versus temperature of $\kappa$-(ET)$_{2}$ABr at temperatures from 2 K to 300 K. The resistivity data are normalized by the one measured at 300 K. The inset shows the enlargement of the data at low temperatures. The arrow points to $T_{c}^{onset} $ of 12.2 K.}
\end{figure}

\begin{figure*}
\centering
\includegraphics[width=16cm]{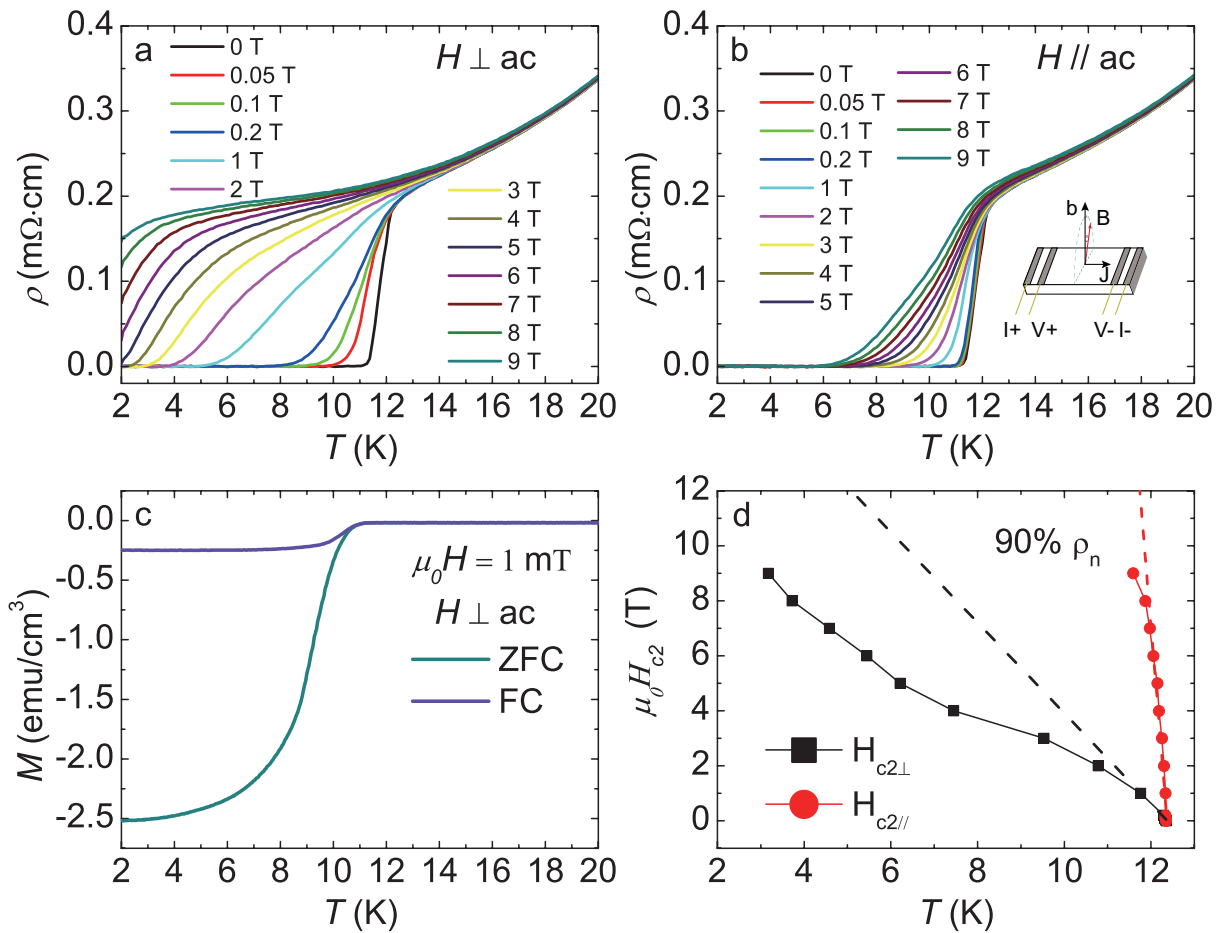}\\[5pt]
\caption{\label{fig2}Temperature dependence of in-plane resistivity of $\kappa$-(ET)$_{2}$ABr under (a) $H \perp ac$ and (b) $H \parallel ac$ between 2 K and 20 K. The applied magnetic field ranges from 0 T to 9 T. (c) Temperature dependence of magnetic susceptibility of $\kappa$-(ET)$_{2}$ABr between 2 K and 20 K. The green and blue curves represent the magnetization measured under zero field cooling (ZFC) and field cooling (FC) modes, respectively. (d) The upper critical field of $\kappa$-(ET)$_{2}$ABr with $H \perp ac$ and $H \parallel ac$, respectively. The dashed lines represent the slopes of $\mu _{0}H_{c2}(T)$ near $T_{c} $.}
\end{figure*}

Fig.~\ref{fig1} shows the normalized in-plane resistivity $\rho(T)/\rho$(300 K) of $\kappa$-(ET)$_{2}$ABr at temperatures from 2 K to 300 K at ambient pressure. The normalized $\rho(T)$ curve displays four distinct subdivisions, which is in agreement with the previous reports\cite{PhysRevLett.104.217003, PhysRevB.90.195150, PhysRevLett.114.216403, PhysRevB.60.9309, kamiya2002magnetic, kund1993anomalous,watanabe1991lattice, yu1991anisotropic}. The arrow in the inset points to the onset of $T_{c} $ ($T_{c}^{onset} $) of 12.2 K.  Below $T_{c}^{onset}$, $\kappa$-(ET)$_{2}$ABr is in the superconducting state. The transition width measured from 90$\%$ to 1$\%$ $\rho_n$ is only about 0.4 K. The resistivity turns into the behavior of $\rho =\rho _{0} +AT^{2}$ in the temperature region $T_{c}^{onset} < T < $ 30 K, which is consistent with the expectation of the Fermi liquid theory for a metal. As temperature increases further, the normalized $\rho(T)$ curve starts to deviate from the $T^{2}$ dependence behavior and displays a dramatic increase followed by a pronounced maximum around 100 K. In the high-temperature region $T >$ 100 K, the resistivity decreases with increasing temperature, showing a semiconducting behavior. This rapid rising of resistivity above about 40 K and the metal-insulator transition was reported in previous literatures\cite{strack2005resistivity, su1998structural, PhysRevB.60.574} and was attributed to the temperature induced evolution of the correlation effect and density of states near Fermi energy\cite{PRLLime}.

\subsection{Temperature dependent anisotropy}

\begin{figure*}
\centering
\includegraphics[width=18cm]{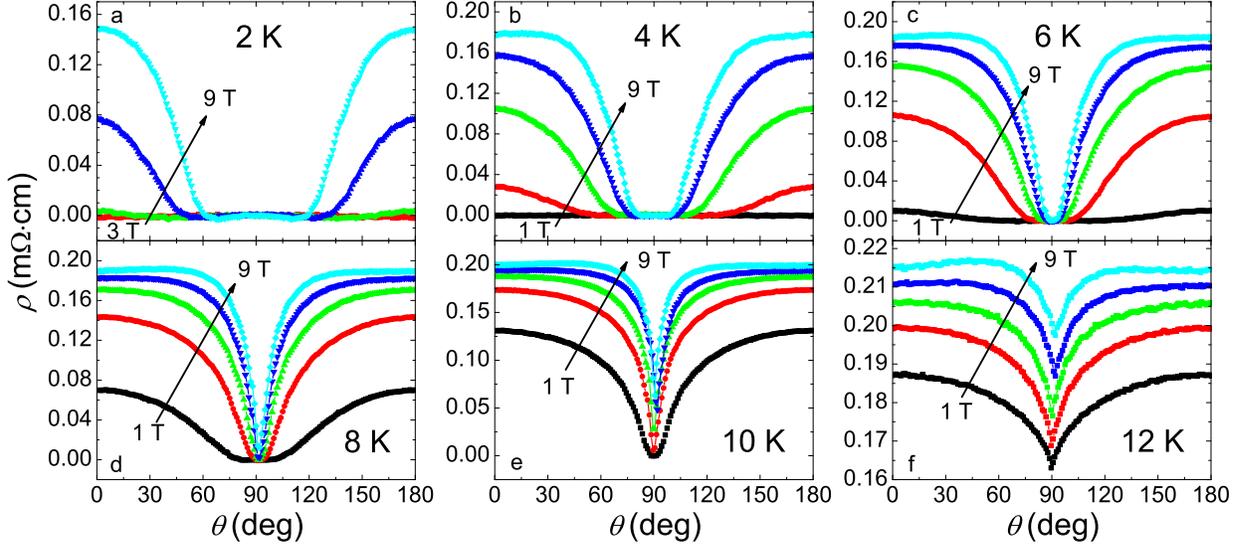}\\[5pt]
\caption{\label{fig3}Angular dependence of in-plane resistivity of $\kappa$-(ET)$_{2}$ABr at  2, 4, 6, 8, 10 and 12 K. (a) The measured magnetic fields are $\mu _{0} H$= 3, 5, 7, 9 T. (b)-(f) The measured magnetic fields are $\mu _{0} H$= 1, 3, 5, 7, 9 T. }
\end{figure*}

In order to intuitively demonstrate the anisotropy of this layered organic superconductor, the temperature dependence of in-plane resistivity was measured with the magnetic field oriented perpendicular ($H \perp ac$) and parallel ($H \parallel ac$) to the ac-planes. The temperature dependence of magnetic susceptibility measurement of the same single crystal was carried out before measuring the resistivity. As shown in Fig.~\ref{fig2}(c), the magnetic susceptibility curves show the significant diamagnetism signal and a sharp superconducting transition, indicating that the single crystal used for the measurements is of high quality. In the configuration of $H \perp ac$ shown in Fig.~\ref{fig2}(a), the superconducting transition widens significantly as the magnetic field increases. However, when $H \parallel ac$, the field induced broadening becomes very weak, the sample $\kappa$-(ET)$_{2}$ABr is still in superconducting state at $\mu _{0}H_{\parallel }$ = 9 T. The slight broadening of the superconducting transition indicates that the parallel upper critical field is quite high. The upper critical field $\mu _{0} H_{c2}$ is then determined by using a criterion of 90$\%\rho _{n}(T)$, where $\rho _{n}(T)$ is determined by a fit to the quadratic temperature dependence $\rho =\rho _{0} +AT^{2}$ in the normal state resistivity between 13 K and 20 K. The field-temperature phase diagram $\mu _{0} H_{c2 }-T $ is depicted in Fig.~\ref{fig2}(d). The slopes of $\mu _{0} H_{c2 } -T $ near $T_{c} $ are -1.65 T/K and -19.2 T/K for $H \perp ac$ and $H \parallel ac$, respectively. According to the Werthamer-Helfand-Hohenberg (WHH) formula\cite{brison1995anisotropy}
\begin{equation}
H_{c2}^{orb}(0)=-0.69(dH_{c2}/dT )\mid_{T_{c} } T_{c},
\label{eq1}
\end{equation}
the extrapolated orbital critical fields are $\mu _{0}   H_{c2} ^{\perp } (0)=13.9$ T and $\mu _{0}   H_{c2} ^{\parallel } (0)=161.6$ T. The coherence length $\xi $ is given by $H_{c2} =\Phi _{0} /2\pi\xi ^{2} $, where $\Phi _{0}$ is the flux quantum. Therefore, it is estimated that the in-plane coherence length $\xi _{\parallel} (0)$ is 4.9 nm and the out-of-plane coherence length $\xi _{\perp} (0)$ is 1.4 nm. The anisotropy $\Gamma$ derived from the formula of $\Gamma =H_{c2}^{\parallel}(0) /H_{c2}^{\perp}(0)$ is 11.6, which indicates highly anisotropic feature of the superconducting state in this organic superconductor.

\begin{figure*}
\centering
\includegraphics[width=16cm]{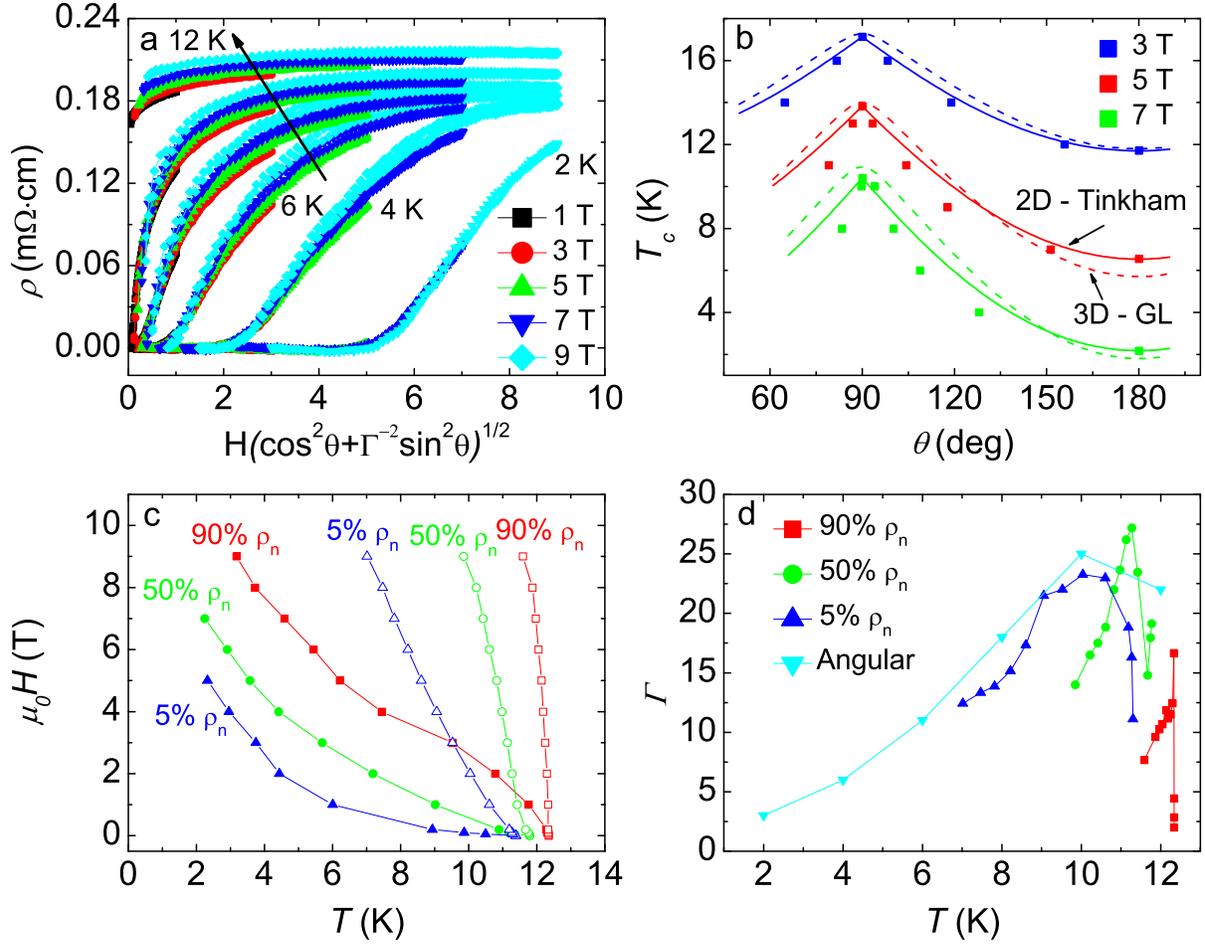}\\[5pt]
\caption{\label{fig4}(a) Scaling of $\rho$ versus scaling variable $\tilde{H}=H\sqrt{cos^{2}\theta+\Gamma ^{-2}sin^{2}\theta} $ based on the anisotropic GL theory. (b) Angle dependence of $T_{c}$ under different magnetic fields ($\mu _{0}H$ =3, 5 and 7 T). The scatter plot shows $T_{c} (\theta )$ extracted from the experimental data. The solid curves and the dashed curves are the fits of $T_{c}(\theta)$ by the 2D-Tinkham model and anisotropic 3D-GL model, respectively. The curves under different magnetic fields are offset by 3 K for clarity. (c) Temperature dependence of critical fields determined by different criteria. The full symbols and the open symbols represent the critical fields perpendicular and parallel to the ac-planes, respectively. The red, green and blue symbols represent critical fields determined using the criteria of $90\%\rho_{n}(T)$, $50\%\rho_{n}(T)$ and $5\%\rho_{n}(T)$, respectively. (d) Temperature dependence of anisotropy $\Gamma$. The red, green and blue symbols are the anisotropy calculated from the formula of $\Gamma =H^{\parallel } /H^{\perp }$ using the data in (c). The cyan symbols are the appropriate $\Gamma$ selected for the scaling in (a).}
\end{figure*}

In fact, the anisotropy $\Gamma$ is not a fixed value at different temperatures. Therefore, exploring the temperature dependence of $\Gamma$ is necessary for a thorough study of the anisotropic properties of this layered organic superconductor. For anisotropic materials, according to the anisotropic GL theory, the angular dependence of $H_{c2}$ can be obtained by following formula
\begin{equation}
H_{c2}^{GL}(\theta ) = \frac{H_{c2}^{\perp }}{(cos^{2}\theta+\Gamma ^{-2}sin^{2}\theta   )^{1/2} }.
\label{eq2}
\end{equation}
By selecting an appropriate value for $\Gamma$, the angle-resolved in-plane resistivity measured under different fields at a fixed temperature can be scaled onto one curve by using the scaling variable $\tilde{H}=H\sqrt{cos^{2}\theta+\Gamma ^{-2}sin^{2}\theta} $\cite{PhysRevLett.68.875}. Fig.~\ref{fig3} displays the angle-resolved in-plane resistivity at six measured temperatures. Here $\theta =0^{\circ} $ and $\theta =180^{\circ} $ correspond to the cases with magnetic field perpendicular to the ac-planes, and $\theta =90^{\circ} $ corresponds to the magnetic field parallel to the ac-planes. The scaling results of $\rho$ versus $\tilde{H}$ are shown in Fig.~\ref{fig4}(a), and the appropriate value of $\Gamma$ selected for each temperature are shown in Fig.~\ref{fig4}(d). One can see that the resistivity curves show good scaling behavior at 2 K and 4 K. However, at higher temperatures, the scaling is not very successful. We extracted $T_{c}$ at different angles $\theta$ under fixed magnetic fields, and adopted the two theoretical models to fit the angle dependence of $T_{c}$, which are shown in Fig. \ref{fig4}(b). The solid line is the fit of $T_{c}(\theta)$ by the 2D-Tinkham model\cite{PhysRevB.40.5263, tinkham1996introduction}, which is described by,
\begin{multline}
T_{c}(\theta ) =T_{c0}-\left | (T_{c0}-T_{c}^{\perp  } (H))cos\theta  \right | \\
 -(T_{c0}-T_{c}^{\parallel } (H))sin^{2}\theta  .
\label{eq3}
\end{multline}
Here, $T_{c0}$ is the superconducting transition temperature in zero-field. $T_{c}^{\perp  } (H)$ and $T_{c}^{\parallel } (H)$ are the superconducting transition temperatures for $H \perp ac$ and $H \parallel ac$, respectively. The dashed line is the fit of $T_{c}(\theta)$ based on anisotropic 3D-GL theory\cite{PhysRevB.40.5263}, which is given by,
\begin{multline}
T_{c} (\theta )=T_{c0}+H_{0}/(\partial H_{c2}^{\perp }/\partial T  )  \\
\times (cos^{2}\theta +m_{\parallel}/m_{\perp }sin^{2}\theta     )^{1/2}.
\end{multline}
Here $H_{0}$ is the applied magnetic field. The ratio of the effective masses is calculated from the formula of $m_{\parallel}/m_{\perp }=(H_{c2}^{\perp}/ H_{c2}^{\parallel})^{2} $. It can be seen from our model that the 2D-Tinkham model fits better than that of the anisotropic 3D-GL model. Fig.~\ref{fig4}(c) displays the temperature dependence of critical fields at both magnetic field directions defined by three different criteria $90\%\rho_{n}(T)$, $50\%\rho_{n}(T)$ and $5\%\rho_{n}(T)$. The anisotropy $\Gamma$ determined by the ratio of parallel critical field to perpendicular critical field is shown in Fig.~\ref{fig4}(d). It is noteworthy that $\Gamma$ defined by the criterion of $5\%\rho_{n}(T)$ has approximately the same temperature dependence as the value determined from the scaling of angular resistivity versus $\tilde{H}$. The anisotropy $\Gamma$ changes a lot from above 20 near $T_{c}$ to a small value at low temperatures, indicating a crossover from the orbital depairing mechanism in high-temperature and low-field region to the paramagnetic depairing mechanism in the high-field and low-temperature region.

\begin{figure*}
\centering
\includegraphics[width=16cm]{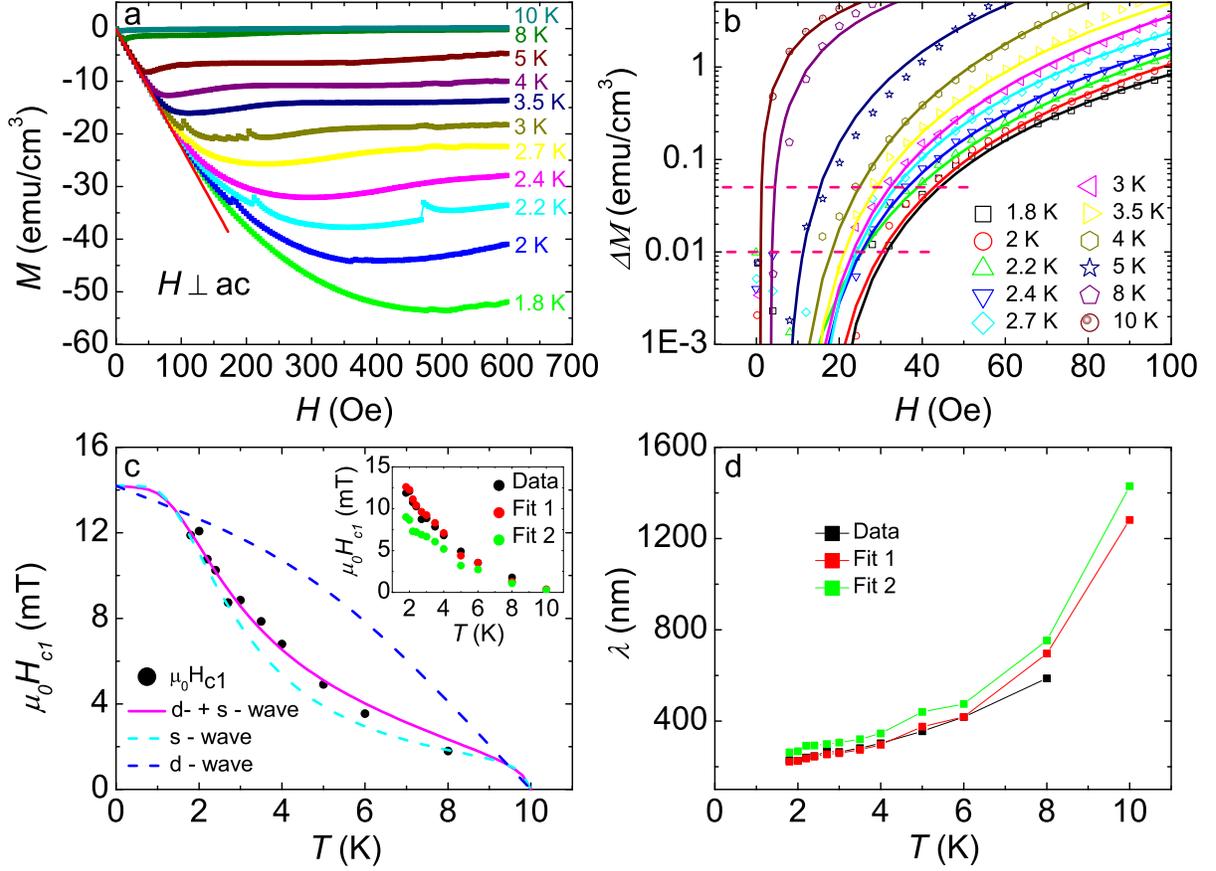}
\caption{\label{fig5}(a) Magnetization versus magnetic field $M(H)$ measured between 1.8 K and 10 K. The applied magnetic field is perpendicular to ac-planes. The straight line represents the magnetization of the Meissner shielding state. (b) The open symbols represent the differences $\bigtriangleup M$ of subtracting the magnetization data of the Meissner state from the magnetization data at each temperature. The solid lines with the same corresponding color represent the magnetization curves fitted using $\bigtriangleup M=a(H-b)^{c} $ at various temperatures and the horizontal dashed lines represent the criteria of 0.05 $\rm emu/cm^{3}$ and 0.01 $\rm emu/cm^{3}$ from top to bottom, respectively. Black symbols in (c) and inset are the temperature dependence of lower critical field $H_{c1}$ extracted from the experimental data defined using the criterion of 0.05 $\rm emu/cm^{3}$. The solid line is the fitting curve obtained by using a two-gap model, and the two dashed lines represent the contributions of the $s$-wave gap and the $d$-wave gap in the two-gap model. The inset also shows two $H_{c1}(T)$ data determined from the fitting data using different criteria $\bigtriangleup M= 0.05$ $\rm emu/cm^{3}$ (Fit 1) and $\bigtriangleup M= 0.01$ $\rm emu/cm^{3}$ (Fit 2). (d) Temperature dependence of penetration depth $\lambda$ calculated from the data of $H_{c1}(T)$.}
\end{figure*}

\subsection{Magnetic field penetration and lower critical field}

The low-field magnetization was measured to explore the magnetic field penetration behavior of $\kappa$-(ET)$_{2}$ABr. Now the magnetic field was applied perpendicular to the ac-planes. The initial magnetization curves in Fig.~\ref{fig5}(a) exhibit a linear behavior, which shows the Meissner state at low magnetic fields. As the applied magnetic field increases further, the diamagnetization curves deviate from linearity, indicating that the magnetic field starts to penetrate into the sample in the form of vortices. The slope of the magnetization curve of the perfect Meissner state should be $-4\pi M/H=-1$, while the slope of the initial linear part in Fig.~\ref{fig5}(a) is -2.82, which indicates that the influence of the demagnetization factor $N$ cannot be ignored. Considering the influence of $N$, the magnetization relationship in the Meissner state is $-4\pi M=H/(1-N) $, where $N$ takes 0.65. The magnetic field that deviates from the linear Meissner shielding of the magnetization curve is defined as the lower critical field $H_{c1}$. The open symbols in Fig.~\ref{fig5}(b) show the deviations of the experimental magnetization data from that of the Meissner state: $\bigtriangleup M=M-M_{Meissner}$. Due to the small value of $H_{c1}$ for $\kappa$-(ET)$_{2}$ABr, especially at a temperature close to $T_{c}$, only experimental data within the measurement accuracy range of SQUID can be provided. Therefore, in order to determine a more accurate $H_{c1}$ by using a lower criterion, we used the relationship $\bigtriangleup M=a(H-b)^{c} $ to fit the experimental $\bigtriangleup M(H)$ curves in the low-field region, which are shown in the form of solid curves in Fig.~\ref{fig5}(b). The fitting parameters at different temperatures are listed in Table~\ref{tab1}. With $\triangle M=0.05$ $\rm emu/cm^{3}$ as the criterion, we also extracted the lower critical fields $H_{c1}(T)$ from the experimental data, the results are shown in Fig.~\ref{fig5}(c). The temperature dependence of $H_{c1}$ extracted from the fitting curves of $\bigtriangleup M(H)$ are shown in the inset, where the determination criterion of Fit 1 (red symbols) is $\triangle M$ = 0.05 $\rm emu/cm^{3}$ and that of Fit 2 (green symbols) is $\triangle M$ = 0.01 $\rm emu/cm^{3}$. The final values of $H_{c1}$ are obtained by multiplying the deviation field data with $1/(1-N)$. All the $H_{c1}(T)$ curves show the concave shape in wide temperature region.

For a single-band superconductor, the relationship between $H_{c1}$ and the normalized superfluid density $\widetilde{\rho } _{s} $ is given by\cite{CARRINGTON2003205, PhysRevLett.94.027001},
\begin{multline}
\frac{H_{c1}(T) }{H_{c1}(0)} =\widetilde{\rho } _{s}(T)=\frac{\rho _{s}(T) }{\rho _{s}(0)}\\
=1+2\int_{0}^{\infty } \frac{\mathrm{d} f(E,T)}{\mathrm{d} E}  \frac{E}{\sqrt{E^{2}-\bigtriangleup (T)^{2}  } } dE.
\end{multline}
Here, $H_{c1}(0)$ and $\rho _{s}(0)$ are the lower critical field and superfluid density in zero-temperature limit, $\bigtriangleup (T)$ is the superconducting gap function, $f(E,T)$ is the Fermi function, and $E=\sqrt{\epsilon ^{2}+\bigtriangleup ^{2}  }$ is the total energy, where $\epsilon$ is the single-particle energy counting from the Fermi energy. As shown by the fitting curves, neither a single $s$-wave gap nor a single $d$-wave gap can fit $H_{c1}(T)$ well over a wide temperature region. However, $H_{c1}(T)$ can be well fitted using a two-gap model containing an $s$-wave gap and a $d$-wave gap. The normalized superfluid density $\widetilde{\rho } _{s}$ of the two-gap model can be expressed as $\widetilde{\rho } _{s} =x\widetilde{\rho } _{s} ^{s} +(1-x)\widetilde{\rho } _{s} ^{d}$, where x is the proportion of $\widetilde{\rho } _{s}$ from the $s$-wave gap. The gap values we use to fit $H_{c1}(T)$ are $\bigtriangleup _{s}=0.5$ meV and $\bigtriangleup _{d}=2.2$ meV. The proportion x of the $s$-wave gap is 0.78. The value of $\bigtriangleup _{d}$ used in the above fitting is consistent with the gap value observed in the STM experiment\cite{doi:10.1143/JPSJ.77.114707}. The existence of $d$-wave component indicates a sign change of the superconducting gap cross the Fermi surface, implying a strong coupling superconductivity in this organic superconductor, and the gap ratio of $2\bigtriangleup _{d}/k_{B}T_{c}$ = 4.18 being larger than the expected value (3.53) of the weak-coupling BCS theory. Similar to our fitting result, $Pinteri\acute{c} $ et al. used an $s+d$-wave gap to well describe the temperature dependence of superfluid density obtained by the ac susceptibility technique\cite{refId0}. In their two-gap model, the proportion of the $s$-wave component is also quite large (about 0.7). Some theoretical calculations point to an eight-node superconductivity with a pairing mechanism of $s_{\pm } +d_{x^{2}-y^{2}  } $-wave symmetry in $\kappa$-(ET)$_{2}$ABr\cite{PhysRevB.97.014530, PhysRevLett.116.237001}, our experimental results for the necessity of nodal gaps may partially support to this theoretical prediction. In the $\kappa$-(ET)$_2$X superconductors, it was predicted that the Fermi surface splits into two parts, one is open (electron-like) and 1D running down two of the Brillouin-zone edges, and another one is a closed `quasi-two-dimensional' hole pocket\cite{singleton2001band}. Thus the discovery of two gaps and two bands from our experiment is understandable. But a momentum resolved gap structure is highly desired. Anyway, the pairing symmetry of $\kappa$-(ET)$_{2}$X is still under debate, and more experimental and theoretical research are needed to provide more explicit conclusions. In Fig.~\ref{fig5}(d), we present the penetration depth $\lambda $ calculated by $\lambda =\sqrt{\frac{\Phi _{0}ln\kappa  }{4\pi H_{c1} } }$, where $\kappa$ is the GL parameter. The penetration depth $\lambda $ shows smooth temperature dependence in the low-temperature region, and it increases sharply at a temperature close to $T_{c} $. However, we need to emphasize that some previous reports concluded that $\lambda$ in zero-temperature limit is about 500-800 nm\cite{PhysRevLett.83.4172, PhysRevResearch.2.043008, doi:10.7566/JPSJ.82.064711}, reflecting a much small superfluid density. In the wide temperature region, the overall values of $\lambda $ we obtained are smaller than the values reported in the literature. It was reported once that the penetration depth at zero temperature is even smaller than the value we obtained\cite{CARRINGTON2003205}. We interpret this discrepancy as a consequence of different sample status from different groups. This is reasonable since the effective DOS near the Fermi energy is heavily influenced by the very narrow and shallow bands crossing the Fermi energy. A slight change of the sample status may greatly modify the superfluid density of the system. Our latest experiments and refined analysis reveal a relatively small penetration depth $\lambda(0)\sim$ 200 nm, implying a moderate superfluid density in the system.

\begin{table}[t]
\caption{\label{tab1}Values of the fitting parameters a, b and c for different temperatures.}
\begin{ruledtabular}
\begin{tabular}{cccc}
\textrm{Temperature (K)}&
\textrm{a}&
\multicolumn{1}{c}{\textrm{b}}&
\textrm{c}\\
\colrule
1.8 & 1.8E-5 & 19.0 & 2.5\\
2 & 4.1E-6 & 14.0 & 2.8\\
2.2 & 6.3E-7 & 5.0 & 3.2\\
2.4 & 6.5E-6 & 11.0 & 2.8\\
2.7 & 1.4E-5 & 12.6 & 2.7\\
3 & 5.0E-6 & 10.6 & 3.0\\
3.5 & 1.1E-5 & 10.0 & 2.9\\
4 & 5.5E-6 & 7.5 & 3.2\\
5 & 2.9E-4 & 7.0 & 2.4\\
8 & 5.1E-2 & 3.5 & 1.3\\
10 & 1.9E-1 & 1.0 & 1.0\\
\end{tabular}
\end{ruledtabular}
\end{table}

\section{CONCLUSION}
In summary, we have conducted a comprehensive analysis of the angle-dependent resistivity and the magnetic field penetration of the layered organic superconductor $\kappa$-(BEDT-TTF)$_{2}$Cu[N(CN)$_{2}$]Br. The anisotropy is determined not only through the comparison of the upper critical fields between the configurations of $H\perp ac$ and $H\parallel ac$, but also through the scaling of the angular dependence of resistivity based on the anisotropic GL-model. It is concluded that the anisotropy $\Gamma $ of $\kappa$-(ET)$_{2}$ABr is strongly temperature dependent. The value of $\Gamma $ near $T_{c} $ is quite large, indicating that $\kappa$-(ET)$_{2}$ABr has an obvious 2D feature. As the temperature decreases, the value of $\Gamma $ gradually changes to a small value, which can be interpreted as a crossover from the orbital depairing mechanism in high-temperature and low-field region to the paramagnetic depairing mechanism in the high-field and low-temperature region. In the configuration of $H\perp ac$, the temperature dependence of lower critical field $H_{c1} $ is obtained from the data of the local magnetization measurement. In a wide temperature region, $H_{c1} (T)$ cannot be fitted with a single gap model. Instead, a two-gap model containing an $s$-wave gap and a $d$-wave gap is used to fit data of $H_{c1} (T)$ well, indicating the multi-gap and unconventional superconductivity in $\kappa$-(ET)$_{2}$ABr. Our experimental results are of great significance for figuring out the unconventional superconducting mechanism of $\kappa$-(ET)$_{2}$ABr.

\begin{acknowledgments}
This work was supported by the National Natural Science Foundation of China (Grant Nos. 11927809, NSFC-DFG12061131001), the state key project (Grant No. 2022YFA1403200) and the Strategic Priority Research Program of Chinese Academy of Sciences (Grant No. XDB25000000).
\end{acknowledgments}

\end{document}